\documentclass{pasj00}

\begin{document}
\SetRunningHead{Y. Takeda and S. Honda}{Oxygen abundance determination of B-type Stars}
\Received{2016/01/07}
\Accepted{2016/02/15}

\title{Oxygen abundance determination of B-type stars \\
with the O~{\sc i} 7771--5~$\rm\AA$ lines
\thanks{Based on data collected by using the NAYUTA Telescope
of the Nishi-Harima Astronomical Observatory.}
}

%

\author{
Yoichi \textsc{Takeda}\altaffilmark{1,2} and
Satoshi \textsc{Honda}\altaffilmark{3}
}
\altaffiltext{1}{National Astronomical Observatory, 2-21-1 Osawa, 
Mitaka, Tokyo 181-8588}
\email{takeda.yoichi@nao.ac.jp}
\altaffiltext{2}{SOKENDAI, The Graduate University for Advanced Studies, 
2-21-1 Osawa, Mitaka, Tokyo 181-8588}
\altaffiltext{3}{Nishi-Harima Astronomical Observatory, Center for Astronomy,\\
University of Hyogo, 407-2 Nishigaichi, Sayo-cho, Sayo, Hyogo 679-5313}

%


\maketitle

\begin{abstract}
Oxygen abundances of 34 B-type stars in the effective temperature range of
$T_{\rm eff} \sim$~10000--28000~K with diversified rotational velocities 
($v_{\rm e}\sin i \sim $~0--250~km~s$^{-1}$) were determined from 
the O~{\sc i} triplet lines at 7771--5~$\rm\AA$, 
with an aim to examine whether this O~{\sc i} feature can be a reliable abundance 
indicator for such high-temperature stars including rapid rotators. 
It revealed that the required non-LTE abundance correction is distinctly
large (ranging from $\sim -0.6$~dex to $\sim -1.7$~dex) and its consideration is 
indispensable. On the condition that the non-LTE effect is taken into account,
this triplet is a useful O abundance indicator (with a precision of $\ltsim 0.2$~dex) 
up to $T_{\rm eff} \ltsim 25000$~K, since its total equivalent width is sufficiently 
large ($\gtsim 200$~m$\rm\AA$). 
In contrast, it is not adequate for abundance derivation for stars at 
$T_{\rm eff} \gtsim 25000$~K, where its strength rapidly drops down
toward a hardly detectable level (except for sharp-lined stars)
and its sensitivity to $T_{\rm eff}$ or $\log g$ becomes considerably large.
The resulting non-LTE oxygen abundances turned out to be 
almost normal (i.e., near-solar around $\sim$~8.7--8.8 within 
$\sim \pm$~0.2~dex) for most stars without any dependence upon 
projected rotational velocity as well as luminosity (or mass),
which is consistent with the prediction of recent stellar evolution calculations.

\end{abstract}

%


\section{Introduction}

The O~{\sc i} triplet lines at 7771--5~$\rm\AA$ 
(3s~$^{5}$S$^{\rm o}$--3p~$^{5}$P, multiplet 1) are known to play 
important roles in stellar spectroscopy, since they are sufficiently strong to be 
visible in stars of wide temperature range (spectral types from K to B) and provide 
us with possibilities of studying not only oxygen abundances but also line-broadening
parameters (such as turbulences or projected rotational velocities). 
Making use of this merit, our group has been exploited these lines for 
investigating the behavior of microturbulent velocities in A--F stars (Takeda 1992)
as well as the O abundances of A--F dwarfs (Takeda \& Sadakane 1997; 
Takeda et al. 2008), G--K giants (Takeda, Kawanomoto, \& Sadakane 1998; 
Takeda et al. 2015), and M~67 stars from the 
turn-off point through the red-giant branch (Takeda \& Honda 2015). 
This time we pay our attention to B-type stars, while expecting a usefulness of 
applying this O~{\sc i} triplet also to this high-temperature regime.

Studying surface oxygen abundances in B-type stars (which have comparatively high 
$T_{\rm eff}$ and $M$) is especially significant in connection with rapid 
rotators, since a deep mixing caused by the meridional circulation may salvage 
ON-cycled material from the interior to produce more or less an underabundance 
of O at the surface, similarly to an overabundance of He due to mixing of 
H$\rightarrow$He processed product. Since rapidly-rotating B-type stars showing
He-enhancement are actually observed (e.g., Lyubimkov et al. 2004), O-deficiency 
may be another indicator for the existence of efficient rotation-induced mixing, 
and thus observational information on the oxygen abundances of a large number of B-type 
stars with various rotational velocities can make an important touchstone for 
the prediction from the stellar evolution theory (e.g., Georgy et al. 2013).

On the observational side, oxygen abundances of B-type stars have so far been studied 
mainly by using O~{\sc ii} lines in the blue region (see, e.g., Gies \& Lambert 1992; 
Kilian 1992; Cunha \& Lambert 1994; Korotin, Andrievsky, \& Luck 1999; Sim\'{o}n-D\'{\i}az 2010; 
Nieva \& Przybilla 2012; Lyubimkov et al. 2013), Although these ionized oxygen lines 
are useful because they are numerous in number, their application is exclusively 
limited to early B stars, since the strengths of these lines quickly fall off 
below $T_{\rm eff} \sim 20000$~K. This is a shortcoming from a viewpoint of 
continuation/comparison with lower temperature stars (A, F, and G types),
for which neutral O~{\sc i} lines have to be used.  Besides, since these previous 
studies using  O~{\sc ii} lines are mostly directed to sharp-lined B stars, prospect 
for applying them to rapid rotators is not clear. 

Meanwhile, spectroscopic investigations on the O abundances of B-type stars
using neutral oxygen lines have been less common and tend to be biased to 
either late-B stars or slow rotators.
Takeda et al. (2010; hereinafter referred to as Paper~I) carried out a non-LTE analysis 
on the O (and Ne) abundances of sharp-lined B stars ($v_{\rm e}\sin i \sim$~0--30~km~s$^{-1}$ 
and 10000~K~$\ltsim T_{\rm eff} \ltsim 23000$~K) by using O~{\sc i} 6156--8 lines 
(multiplet 10), which however are inapplicable to rapid rotators because of their weakness.
Niemczura, Morel, and Aerts's (2009) LTE study on the abundances of oxygen (along with 
other elements) based on weaker O~{\sc i} lines\footnote{Although they did not explicitly 
publish the list of used lines, they must have exploited weaker O~{\sc i} lines 
(presumably those at 6156--8~$\rm\AA$), since they stated that O~{\sc i} 7771--5 lines
were excluded because of the strong non-LTE effect.} is limited to B6--B9.5 stars.
Similarly,  only late-B (B5--B9) stars were targeted in Hempel and Holweger's (2003) 
determination of non-LTE oxygen abundances based on O~{\sc i} 7771-5 lines.

Among the previous abundance studies of B-type stars, most notable 
in connection with our present interest would be the recent one of Nieva 
and Przybilla (2012), who determined the abundance of O (along with 
those of He, C, N, Ne, Mg, Si, and Fe) for 20 sharp-lined mid-to-early 
B stars (16000~K $\ltsim T_{\rm eff} \ltsim$ 33000~K) by using 
the state-of-the-art non-LTE technique based on high-S/N and 
high-resolution spectra. While their analysis mainly depends on 
a number of O~{\sc ii} lines, they also employed O~{\sc i} lines 
(e,g.,  O~{\sc i} features at 6156--8~$\rm\AA$, 7771--5~$\rm\AA$, 
and 8446~$\rm\AA$) for stars with $T_{\rm eff} \ltsim 27000$~K, 
in order to check that O~{\sc i}/O~{\sc ii} ionization equilibrium 
is accomplished. Although their work is regarded as an important benchmark, 
only apparently slow rotators ($v_{\rm e}\sin i \ltsim 30$~km~s$^{-1}$) 
were treated and rapid rotators were out of their consideration.

This situation motivated us to investigate a possibility of applying 
O~{\sc i} 7771--5 lines to B stars with much wider range stellar properties; i.e., 
$T_{\rm eff} \sim$~10000--30000~K corresponding to late B through early B type, 
slow as well as rapid rotators covering $0 \ltsim v_{\rm e}\sin i \ltsim$~200--300~km~$^{-1}$.\\
--- Is this O~{\sc i} triplet feature still usable for early B-type stars 
($T_{\rm eff} \sim$~20000--30000~K) of various rotational velocities, despite 
that its strength should be considerably weakened as most O atoms are ionized at 
such high $T_{\rm eff}$ range? \\
---  How is the behavior of its strength in terms of the stellar parameters; 
e.g., $T_{\rm eff}$, surface gravity ($\log g$), microturbulence ($\xi$), 
and oxygen abundance? Is it suitable for abundance determination?\\
---  How much important is the non-LTE effect regarding this strong O~{\sc i} triplet? 
Does the required non-LTE abundance correction depend on the atmospheric parameters?

Toward answering these questions, we first conducted non-LTE calculations on an extensive 
parameter grid in order to understand how the O~{\sc i} 7771--5 strengths of B stars behave 
in terms of various key parameters, and then carried out an observational study of 
non-LTE oxygen abundance determinations for 34 selected late-B through early-B type stars 
with diversified $v_{\rm e}\sin i$ values based on the spectra newly obtained at 
Nishi-Harima Astronomical Observatory. The purpose of this paper is to report the 
consequence of this investigation.

The remainder of this article is organized as follows:
The observational data of our program stars are described in section~2, 
and their spectroscopic parameters are derived in section~3. 
We explain our non-LTE calculations in section~4, and the procedures of abundance
determination (including profile fitting, equivalent width evaluation, and error 
estimation) in section~5. The resulting behavior of non-LTE correction and the trend 
of oxygen abundances of B-type stars are discussed in section~6, followed by 
section~7 where the conclusions are summarized.

\section{Observational data} 

The list of our 34 targets is presented in table 1, all of which are apparently 
bright ($V \le 5$~mag) and selected from the extensive $v_{\rm e}\sin i$ compilation 
of 1092 northern B-type stars published by Abt, Levato, and Grosso (2002),
where B supergiants and Be stars were excluded from our selection.
Our sample includes stars of diversified $v_{\rm e}\sin i$ values (from slow 
rotators to rapid rotators up to $v_{\rm e}\sin i \sim 200$~km~s$^{-1}$).
These program stars are plotted on the $\log L$ vs. $\log T_{\rm eff}$ 
diagram (theoretical HR diagram) in figure 1, where Girardi et al.'s (2000)
theoretical evolutionary tracks corresponding to different stellar 
masses are also depicted. We can see from this figure that the masses 
of our sample stars are in the range between $\sim 2.5 M_{\odot}$ and 
$\sim 15 M_{\odot}$.

Spectroscopic observations of these stars were carried out in the summer season 
of 2015 (May through September; observation date for each star is given in table~1) 
by using the Medium And Low-dispersion Longslit Spectrograph (MALLS; cf. 
Ozaki \& Tokimasa 2005) installed on the Nasmyth platform of 
the 2~m NAYUTA telescope at Nishi-Harima Astronomical Observatory (NHAO). 
Equipped with a 2K$\times$2 K CCD detector (13.5 $\mu$m pixel), MALLS
can record a spectrum covering $\sim$~400~$\rm\AA$ (7600--8000~$\rm\AA$)
in the medium-resolution mode with the resolving power of $R \sim 12000$. 
While the exposure time of one frame was typically a few minutes, several 
spectral frames were co-added to improve the signal-to-noise ratio.
The reduction of the spectra (bias subtraction, flat-fielding, 
spectrum extraction, wavelength calibration, co-adding of frames to improve S/N, 
continuum normalization) was performed by using the ``noao.onedspec'' package of 
the software IRAF\footnote{IRAF is distributed
    by the National Optical Astronomy Observatories,
    which is operated by the Association of Universities for Research
    in Astronomy, Inc. under cooperative agreement with
    the National Science Foundation.} 
in a standard manner. 
The S/N ratios of the resulting spectra were estimated to be around $\sim$~200--300
in most cases (except for HR~0779, for which S/N is only $\sim$~60--70), which turned out 
to be sufficient for our purposes. Our final spectra in the 7700--7850~$\rm\AA$ 
region for each of the 34 stars are shown in figure~2.

\section{Atmospheric parameters}

The effective temperature and the surface gravity of each program star 
were determined from the colors of 
Str\"{o}mgren's $uvby\beta$ photometric system with the help of 
Napiwotzki, Sc\"{o}nberner, and Wenske's (1993) {\tt uvbybetanew}
program,\footnote{
$\langle$http://www.astro.le.ac.uk/\~{}rn38/uvbybeta.html$\rangle$.} 
where the observational data of
$b-y$, $c_{1}$, $m_{1}$, and $\beta$ were taken from Hauck and 
Mermilliod (1998) via the SIMBAD database.
The resulting $T_{\rm eff}$ and $\log g$ are summarized in table 1.
The model atmosphere for each star was then constructed
by two-dimensionally interpolating Kurucz's (1993) ATLAS9 
model grid in terms of $T_{\rm eff}$ and $\log g$, where
we exclusively applied the solar-metallicity models. 

As to typical errors of these parameters, we may estimate 
$\sim 3\%$ in $T_{\rm eff}$ and $\sim 0.2$~dex in $\log g$ 
for most of our sample stars (i.e., 32 stars at 
10000~K~$\ltsim T_{\rm eff} \ltsim$~24000~K) 
according to Napiwotzki et al. (1993; cf. section 5 therein).
We should note, however, that parameters derived from Napiwotzki et al.'s 
code would not be reliable any more for stars at $T_{\rm eff} > 24000$~K, 
as pointed out by Nieva (2013) based on the results of Nieva and Przybilla
(2012) who used non-LTE ionization equilibria of various species along 
with the profiles of Balmer lines to determine the atmospheric parameters.
Accordingly, the parameters of the relevant two stars (HR~8238 and HR~8797; 
cf. table~1) having particularily high $T_{\rm eff}$ (25231~K and 27853~K, 
respectively) should be viewed with caution and their errors may be larger
than the typical amounts mentioned above. 

Given that HR~779 (HD~16582) and HR~8238 (HD~205021) in our sample
were also studied by Nieva and Przybilla (2012), it is interesting 
to compare our color-based $T_{\rm eff}$ and $\log g$ with their 
spectroscopically determined values which may be more reliable. 
The results of ($T_{\rm eff}^{\rm ours}/T_{\rm eff}^{\rm Nieva}$, 
$\log g^{\rm ours}/\log g^{\rm Nieva}$) are (21747/21250, 3.63/3.80) 
for HR~779 and (25231/27000, 3,63/3.95) for HR~8238.  We can see 
a reasonable consistency within the expected errors for the former 
B2 star ($T_{\rm eff} < 24000$~K), while the discrepancy is larger 
for the latter B0.5 star ($T_{\rm eff} > 24000$~K) which may be 
attributed to the reason mentioned above (i.e., outside of the 
reliability limit of Napiwotzki et al.'s calibration).

Regarding the microturbulence, which is necessary for abundance
determination, we had to assign an appropriate value since it was not 
possible to establish this parameter based on our spectra.
Here, we divide our targets into two classes of $T_{\rm eff} < 15000$~K
and $T_{\rm eff} > 15000$~K, and tentatively assume $\xi = 1(\pm 1)$~km~s$^{-1}$
for the former and $\xi = 3(\pm 2)$~km~s$^{-1}$ for the latter, based on
the discussion in Takeda et al. (2014; cf. section~3 therein) and 
Paper~I (cf. section~3 therein), respectively.
Admittedly, this is a rather rough choice; but it appears to be  
reasonable as will be discussed in subsection~6.2.

\section{Non-LTE calculations} 

In order to evaluate the non-LTE effect on the strength of O~{\sc i} 7771--5 
triplet, we carried out non-LTE calculations for oxygen on an extensive grid 
of solar-metallicity model atmospheres resulting from combinations of eleven 
$T_{\rm eff}$ values (9000, 10000, 12000, 14000, 16000, 18000, 20000, 22000, 
24000, 26000, 28000~K) and four $\log g$ values (3.0, 3.5, 4.0, 4.5),\footnote{
Since ATLAS9 model grid does not include $\log g= 3.0$ models at  
$T_{\rm eff} \ge 27000$~K presumably because of an instability problem,   
models of only three gravities ($\log g$ = 3.5, 4.0, and 4.5) 
could be used for $T_{\rm eff}$ = 28000~K.}
which sufficiently cover the parameter ranges of our program stars.
The statistical equilibrium calculations for oxygen were done
in the same manner as in Paper~I (cf. subsection~4.1 therein).
Our non-LTE calculation program is based on the code described in 
Takeda (1991). The atomic model of oxygen is the same as that adopted by 
Takeda (2003; cf. subsection~2.1 therein), which is based on the atomic 
data given in Kurucz and Bell (1995). Other various fixed collisional
and radiative rates were evaluated as described in Takeda (1992).
The non-LTE departure coefficients [$b(\tau)$]
applied to each star (to be used for non-LTE abundance determination in section~5) 
were then derived by two-dimensionally interpolating this grid in terms of 
$T_{\rm eff}$ and $\log g$, as was done for model atmospheres.

Toward understanding the behavior of the non-LTE effect to be discussed in 
subsection~6.1, theoretical non-LTE and LTE equivalent widths ($W_{\rm N}$ and 
$W_{\rm L}$) of the whole triplet were computed on this grid of models for 
combinations of four oxygen abundances ([O/H]\footnote{[O/H] 
[$\equiv A_{\rm star}$(O) $- A_{\odot}$(O)] is the differential 
logarithmic oxygen abundance relative to the Sun, where $A$ is 
defined in the usual normalization of $A$(H) = 12.00 and we employed
$A_{\odot}$(O) = 8.93 in order to maintain consistency with the 
similar non-LTE grid computed by Takeda (2003) for FGK stars.
Note that this reference solar abundance (8.93), which was taken 
from Anders and Grevesse (1989) and adopted as the standard value 
in the ATLAS9 program, is appreciably higher by Asplund et al.'s 
(2009) recently updated value (8.69) based on their 3D hydrodynamical
solar model photosphere. However, it is closer to the solar non-LTE
oxygen abundance (8.81) derived by Takeda and Honda (2005) with 
the same O~{\sc i} 7771--5 lines (see the discussion in subsection~5.1 
of Paper~I).} = $-0.6$, $-0.3$, 0.0, and +0.3) and three microturbulences
($\xi$ = 1, 3, and 5~km~s$^{-1}$) by using Kurucz's (1993) WIDTH9 program\footnote{
The original WIDTH9 program had been considerably modified in various respects. 
In order to enable simulation of line profiles by taking into account the
non-LTE effect, the LTE line opacity and the LTE line source function  
usually used in the WIDTH9 program were multiplied by the depth-dependent 
NLTE-to-LTE line opacity ratio ($l_{0}^{\rm NLTE}/l_{0}^{\rm LTE}$) and 
the NLTE line source function ratio ($S_{\rm L}/B$) resulting from our non-LTE
calculation code, respectively (cf. subsection 4.2 in Takeda 1991 for more details).
Besides, since the original WIDTH9 program can handle only a single isolated line,
a spectrum synthesis technique was incorporated for calculating the total equivalent
width of a multi-component feature (such as O~{\sc i} 7771--5).} 
with the atomic data presented in table~2, where $W$ was evaluated by directly 
integrating the synthesized profiles of triplet lines over the wavelength region between 
7770.44~$\rm\AA$ and 7776.94~$\rm\AA$ (i.e., 1.5~$\rm\AA$ margin both shortward
of the 7771.94 line and longward of the 7775.39 line).
We also evaluated the corresponding non-LTE abundance corrections ($\Delta$) 
defined as $A_{\rm N}- A_{\rm L}$, where $A_{\rm N}$ and $A_{\rm L}$
are the abundances derived by inversely analyzing $W_{\rm N}$ in non-LTE
and LTE, respectively. The resulting $W$ and $\Delta$ values are summarized
in the electronic data tables (file name: ncor7773\_xi?.dat) available
as on-line materials. The behaviors of the computed $W$ and $\delta$ in terms of
the atmospheric parameters are shown in figure 3 for selected representative cases,
on which we will focus later in subsection~6.1.  

\section{Abundance determination}

As in Paper~I, our strategy of determining the abundance and related quantities 
(e.g., non-LTE correction, uncertainties due to ambiguities of atmospheric 
parameters) consists of the following consecutive steps: 
(i) derivation of provisional abundance solution 
by applying the spectrum-fitting technique, (ii) evaluation of 
the equivalent width ($W$) by using this fitting-based abundance, 
and (iii) analysis of such derived $W$ with the standard parameters
and changing the conditions (perturbed parameters, LTE and non-LTE). 

\subsection{Synthetic spectrum fitting}

We first searched for the solutions for the oxygen abundance ($A$), projected 
rotational velocity ($v_{\rm e}\sin i$), and radial velocity ($V_{\rm rad}$) 
such as those accomplishing the best fit (minimizing $O-C$ residuals) 
between the theoretical and observed spectrum in the $\sim$~7765--7780~$\rm\AA$ 
region, while applying the automatic fitting algorithm (Takeda 1995). 
In this preparatory step, we used the model atmosphere (cf. section~3) and 
the non-LTE departure coefficients (section~4) prepared for each star
but adopted the same microturbulence of $\xi =3$~km~s$^{-1}$  
for computing the non-LTE theoretical spectrum. Regarding the macroscopic 
line-broadening function to be convolved with the intrinsic spectrum,
we included the instrumental broadening (assumed to be the Gaussian function) 
corresponding to $R\sim 12000$ and the rotational broadening with the limb-darkening
coefficient of $\epsilon = 0.3$ (estimated from Fig.~17.6 of Gray 2005)
How the theoretical spectrum for the converged solutions fits well 
with the observed spectrum is demonstrated in figure~4.\footnote{Note that, 
in the evaluation of $O-C$ residuals, we sometimes masked some regions 
showing features irrelevant to stellar spectra, such as 
spurious spectrum defect, as highlighted in green in this figure.}
The resulting $v_{\rm e}\sin i$ values are presented in table 1,
which are also compared with those given in Abt et al.'s (2002) 
catalogue in figure~5.

\subsection{Derivation and analysis of equivalent width}

Next, we inversely computed the equivalent width ($W$) for the whole 
O~{\sc i} 7771--5 triplet (as done in section~4) 
by using the abundance solution (derived from the fitting analysis) 
along with the same model atmosphere/parameters.
Based on such evaluated $W$ value, the non-LTE abundance ($A_{\rm N}$) 
as well as LTE abundance ($A_{\rm L}$) were freshly computed by applying 
the microturbulence assigned to each star (1 or 3~km~s$^{-1}$ depending on 
$T_{\rm eff}$; cf. section~3), from which the non-LTE correction ($\Delta$) 
was further derived. These $W$, $A_{\rm N}$, and $\Delta$ values are also 
given in table 1.

\subsection{Sensitivity to parameter perturbation} 

We also estimated the errors in $A_{\rm N}$ caused by ambiguities 
involved in the standard atmospheric parameters (presented in table~1).
Since their typical uncertainties are $\pm 3\%$ in $T_{\rm eff}$, 
$\pm 0.2$~dex in $\log g$, and $\pm 1$~km~s$^{-1}$ (for $T_{\rm eff} < 15000$~K) 
or $\pm 2$~km~s$^{-1}$ (for $T_{\rm eff} > 15000$~K) in $\xi$ (cf. section 3). 
six kinds of abundance variations ($\delta_{T+}$, $\delta_{T-}$, 
$\delta_{g+}$, $\delta_{g-}$, $\delta_{\xi+}$, and $\delta_{\xi-}$) were 
derived by repeating the analysis on the $W$ values while perturbing 
the standard atmospheric parameters interchangeably by these amounts.
Finally, the root-sum-square of these perturbations,
$\delta_{Tg\xi} \equiv (\delta_{T}^{2} + \delta_{g}^{2} + \delta_{\xi}^{2})^{1/2}$, 
was regarded as the abundance uncertainty (due to combined errors in 
$T_{\rm eff}$, $\log g$, and $\xi$), 
where $\delta_{T}$, $\delta_{g}$, and $\delta_{\xi}$ are defined as
$\delta_{T} \equiv (|\delta_{T+}| + |\delta_{T-}|)/2$, 
$\delta_{g} \equiv (|\delta_{g+}| + |\delta_{g-}|)/2$, 
and $\delta_{\xi} \equiv (|\delta_{\xi+}| + |\delta_{\xi-}|)/2$,
respectively. 
The resulting abundances ($A_{\rm N}$), equivalent widths ($W$), 
non-LTE corrections ($\Delta$), and abundance variations ($\delta$'s) 
in response to parameter changes, are graphically displayed as functions 
of $T_{\rm eff}$ in figure~6, which we will discuss in subsection~6.2. 

\section{Discussion}

\subsection{Behavior of line strength and non-LTE correction}

Before discussing the results of our analysis derived in section~5 for 
the program stars, we first review how the strength of O~{\sc i} 7771--5 
triplet and its non-LTE effect behaves in terms of the atmospheric parameters.
The following characteristics can be summarized by inspection of figure~3:\\
---  Generally, $W$ declines with an increase in $T_{\rm eff}$.
While its change is gradual at $T_{\rm eff} \ltsim 25000$~K, a rather 
abrupt drop of $W$ occurs at $T_{\rm eff} \gtsim 25000$~K where oxygen 
is predominantly ionized and very few neutral oxygen atoms remain.\\
--- Naturally, $W$ increases with an increase in [O/H] and $\xi$, 
though it is almost $\xi$-independent at $T_{\rm eff} \gtsim 25000$~K
where the line is weak in the linear part of the curve of growth.\\
--- The behavior of $W$ in response to changing $\log g$ is somewhat
complicated (figure~3c). While the line strength slightly grows with a lowering 
of $\log g$ at $T_{\rm eff} \ltsim 25000$~K (because of the increased non-LTE effect; 
note that this tendency is seen only in $W_{\rm N}$ but not in $W_{\rm L}$), 
the trend is reversed at $T_{\rm eff} \gtsim 25000$~K
where $W$ markedly weakens with a decrease in $\log g$ because the effect of 
enhanced ionization becomes more important at the condition of lower-density.\\
--- The non-LTE effect always acts in the direction of intensifying
the line strength ($W_{\rm N} > W_{\rm L}$), which makes 
the non-LTE abundance correction ($\Delta$) always negative.\\
--- The extent of $|\Delta|$ has a peak at $T_{\rm eff} \sim$~15000--20000~K 
and tends to be larger for higher [O/H], smaller $\xi$, and lower $\log g$, 
which can be interpreted that the non-LTE effect becomes more significant 
when the line-forming depth is higher or in the lower density region. 

\subsection{Trend of oxygen abundances} 

We now discuss figure~6. The equivalent widths ($W$) derived in 
subsection~5.2 show a decreasing trend with an increase in $T_{\rm eff}$ 
(figure~6a), which is reasonably consistent with the theoretically 
 predicted trend for $W_{\rm N}$ (cf. figure~3a--c). 

The $|\Delta|$ values are in the range of $\sim$~0.6--1.7~dex (figure~6b), 
which are so large that these corrections are indispensable for reliable 
determination of oxygen abundances of B-type stars based on the 
O~{\sc i}~7771--5 lines. Comparing these results with those of Hempel and 
Holweger (2003), who determined the non-LTE oxygen abundances of B5--B9 stars
($T_{\rm eff} \sim$~9000--17000~K) based on the same O~{\sc i} triplet,
we see that their $|\Delta|$ values ($\sim$~0.3--1.2; cf. their table~5) 
are considerably smaller than ours. We suspect that they underestimated
the importance of the non-LTE effect, which may explain the significantly 
supersolar [O/H]$_{\rm NLTE}$ results (by $\sim$~+0.3--0.6~dex) 
they obtained for many of their sample stars.

Figure~6c suggests that the oxygen abundances derived for the 34 B-type 
stars distribute around $\sim$~8.7--8.8 without any systematic
$T_{\rm eff}$-dependence. Actually the average abundance over all 34 stars
makes $\langle A \rangle = 8.72$ (with standard deviation $\sigma =  0.16$).
If three stars with $A$(O)~$< 8.4$ (HR~6779, 7236, 8238) showing appreciable 
deviations are excluded, we have $\langle A \rangle = 8.76$ ($\sigma =  0.10$).
This means that the surface oxygen abundances of these B-type stars 
are mostly near to the solar composition (see footnote~5), which is 
in fairly good agreement with 
the previous studies for B-type stars; e.g., $8.71 \pm 0.06$ derived in 
Paper~I based on O~{\sc i} 6156--8 lines or $8.73 \pm 0.13$ by Lyubimkov et al. 
(2013) based on O~{\sc ii} lines (see also table~6 and table 7 therein).

Regarding the abundance errors caused by uncertainties in atmospheric parameters,
$\delta_{\xi}$ ($\sim$~0.1--0.2~dex) is predominantly important over   
$\delta_{T}$ ($< 0.1$) or $\delta_{g}$ ($< 0.1$) at $T_{\rm eff} \ltsim 25000$~K, 
to which most of the program stars are relevant (cf. figure 6d--f). Accordingly,
the extents of $\delta_{tg\xi}$ (error bars attached to the symbols in figure~6c)
are essentially determined by $\delta_{\xi}$ and on the order of $\sim$~0.1-0.2~dex
(except for the two stars of $T_{\rm eff} \gtsim 25000$~K),
which is consistent with the standard deviation of our abundance results.
It may appropriate here to comment on the assigned values of $\xi$ 
mentioned in section~3. Since the abundances derived for given 
microturbulences of $\xi = 1 (\pm 1)$~km~s$^{-1}$ ($T_{\rm eff} < 15000$~K) 
and $\xi = 3 (\pm 2)$~km~s$^{-1}$ ($T_{\rm eff} > 15000$~K) do not show
any systematic trend in terms of $T_{\rm eff}$, this may imply that our choice 
is reasonably justified. If a constant $\xi$ were assigned
for all stars, we would have obtained a spurious $T_{\rm eff}$-dependence.\footnote{
Actually, the first provisional abundances derived from the spectrum synthesis 
by tentatively using $\xi = 3$~km~s$^{-1}$ (cf. subsection~5.1) showed
a tendency of appreciable underabundance in late B-type stars.}

The oxygen abundances of 34 stars are plotted against $v_{\rm e}\sin i$ 
as well as $\log L$ in figure~7, where we can not see any systematic
dependence. This means that no appreciable O abundance anomaly is detected
in our sample stars at least within the precision of $\ltsim$~0.1--0.2~dex.
According to Georgy et al.'s (2013) calculation, while no change in the 
surface composition is produced at all during the main sequence period
for the case of non-rotating stars, rapidly rotating stars may show 
some abundance peculiarities caused by the 
rotational mixing of nuclear processed product.
For example, regarding the cases of solar-metallicity ($Z = 0.014$) models,
the changes of He$|$O abundances at the end of the main sequence
for 4~$M_{\odot}$ star is $+0.2\%|-0.7\%$ ($\Omega/\Omega_{\rm crit}$ = 0.5)
and $+3.1\%|-6.2\%$ ($\Omega/\Omega_{\rm crit}$ = 0.95), while those
for 9~$M_{\odot}$ star is $+1.9\%|-4.9\%$ ($\Omega/\Omega_{\rm crit}$ = 0.5)
and $+11.4\%|-17.3\%$ ($\Omega/\Omega_{\rm crit}$ = 0.95), where $\Omega_{\rm crit}$
is the angular rotational velocity corresponding to the critical break-up.
However, even if such an abundance change as large as $\sim$~20\% exists, it 
corresponds to only $\ltsim 0.1$~dex in the logarithmic scale. Thus, 
it is no wonder that we could not detect such a subtle amount, given that 
the precision of our abundance determination is $\ltsim 0.2$~dex.
In this sense, we may state that our result is more or less consistent with
the recent stellar evolution calculations based on the canonical mixing theory.

\subsection{Usability of O~I triplet lines for abundance determination}

Finally, we turn to the question which motivated this investigation (cf. section~1): 
Is the O~{\sc i} 7771--5 triplet practically useful for oxygen abundance 
determination of B-type stars? Our answer is as follows:\\
--- Yes, it is surely useful as far as stars of $T_{\rm eff} \ltsim 25000$~K 
are concerned (given that the non-LTE effect is properly taken into account), 
since its strength is still sufficiently large ($W \gtsim 200$~m$\rm\AA$) 
and abundances are not seriously affected by uncertainties in the atmospheric 
parameters.\\ 
--- For example, if we consider a case of 
$v_{\rm e}\sin i \sim 300$~km~s$^{-1}$ and $W \sim 200$~m$\rm\AA$, 
the FWHM and the central depth of the merged triplet feature would 
be $\sim 10$~$\rm\AA$ and $\sim 2$\%, respectively, which is measurable
without difficulty on a spectrum with S/N ratios of a few hundred.\\
--- In contrast, regarding very early B stars of $T_{\rm eff} \gtsim 25000$~K
(corresponding to $\sim$~B0), the strength of this triplet quickly drops away
and its sensitivity to ambiguities of $T_{\rm eff}$ or $\log g$ becomes
seriously large, which makes it unsuitable as a reliable abundance indicator.\\  
--- We should keep in mind, however, even for the well-behaved case of 
$T_{\rm eff} \ltsim 25000$~K stars, the precision of abundance derivation would 
be on the order of $\ltsim 0.2$~dex, which is mainly restricted by uncertainties 
in the choice of microturbulence. Accordingly, very high-precision should not be 
expected.\\
--- In this sense, detection of surface oxygen abundance anomaly based on 
these O~{\sc i} lines would be feasible only if the extent of peculiarity
is sufficiently large (e.g., by $\sim 0.3$~dex), such as the case where 
non-canonical deep mixing is relevant. The slight abundance change
(up to $\ltsim 0.1$~dex) predicted by the current canonical rotational mixing 
theory can hardly be detected.

\section{Summary and conclusion}

In order to examine whether the O~{\sc i} triplet lines at 7771--5~$\rm\AA$ can 
be used as a reliable oxygen abundance indicator for B-type stars of high 
$T_{\rm eff}$ including rapid rotators, we carried out oxygen abundance determinations 
for selected 34 B-type stars based on the spectra ($R\sim 12000$) obtained by 
the 2~m NAYUTA Telescope at Nishi-Harima Astronomical Observatory. 

Regarding the atmospheric parameters, $T_{\rm eff}$ and $\log g$ were 
determined from $uvby\beta$ colors and the microturbulence was assumed 
to be $\xi = 1 (\pm 1)$~km~s$^{-1}$ ($T_{\rm eff} < 15000$~K) 
and $\xi = 3 (\pm 2)$~km~s$^{-1}$ ($T_{\rm eff} > 15000$~K)
by consulting previous studies. The non-LTE effect was properly
taken into consideration based on our non-LTE calculations
carried out for an extensive grid of models.

We first determined the provisional abundance and $v_{\rm e}\sin i$
by applying the synthetic spectrum-fitting analysis to the 7771--5 feature,
and then inversely determined (from the tentative abundance) the 
equivalent width of the whole triplet to be used for further 
evaluations of non-LTE/LTE abundances, non-LTE correction, and 
errors due to ambiguities of atmospheric parameters.  
  
It was found that the extent of the (negative) non-LTE correction is 
appreciably large to be $|\Delta| \sim$~0.6--1.7~dex and its 
consideration is indispensable. The resulting non-LTE oxygen abundances 
turned out to be nearly solar ($\langle A \rangle \sim$~8.7--8.8 with 
$\sigma \ltsim 0.2$~dex) without any clear dependence upon rotation 
as well as luminosity (or mass), which is consistent with the results 
of published observational studies and stellar evolution calculations
with the canonical mixing theory. 

We concluded that this triplet is a useful O abundance indicator (with a 
precision of $\ltsim 0.2$~dex) up to $T_{\rm eff} \ltsim 25000$~K, since 
its total equivalent width is sufficiently large ($\gtsim 200$~m$\rm\AA$). 
In contrast, it is not adequate for abundance determination for stars at 
$T_{\rm eff} \gtsim 25000$~K, where its strength rapidly turns down and 
its sensitivity to $T_{\rm eff}$ or $\log g$ becomes considerably large.

\bigskip

This research has made use of the SIMBAD database, operated by
CDS, Strasbourg, France.

\clearpage
\onecolumn

\begin{figure}
  \begin{center}
    \FigureFile(100mm,100mm){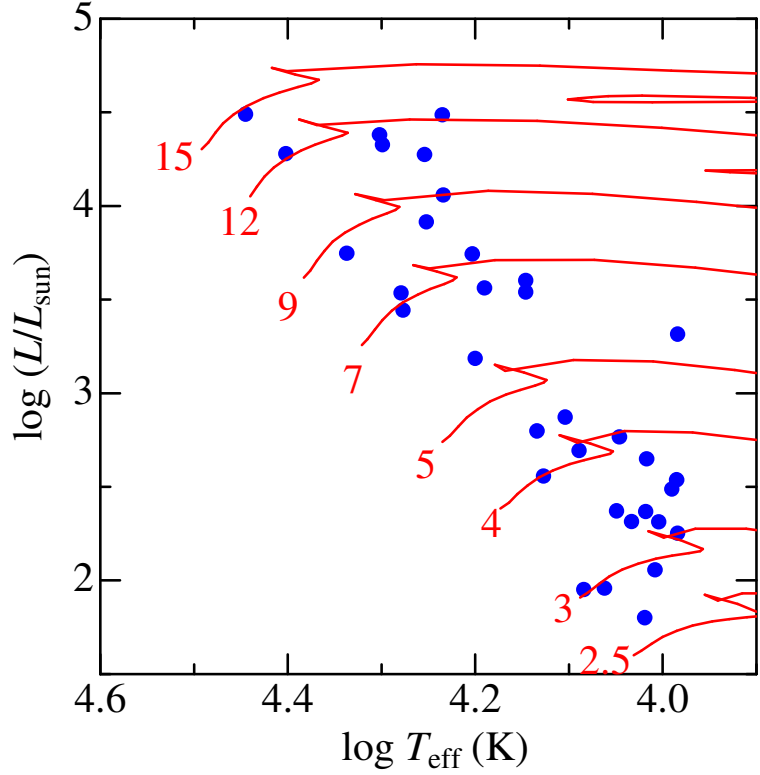}
  \end{center}
\caption{
Plots of the 34 program stars on the theoretical HR diagram
($\log (L/L_{\odot})$ vs. $\log T_{\rm eff}$), where the effective 
temperature ($T_{\rm eff}$) was determined from $uvby\beta$ photometry 
as described in section 3 and the bolometric luminosity ($L$) was 
evaluated from the apparent visual magnitude, Hipparcos parallax 
(ESA 1997), Arenou, Grenon, and G\'{o}mez's (1992) interstellar extinction
map, and Flower's (1996) bolometric correction as in Paper~I. 
Theoretical evolutionary tracks corresponding to the solar metallicity 
computed by Girardi et al. (2000) for eight different initial masses 
(2.5, 3, 4, 5, 7, 9, 12, and 15~$M_{\odot}$) are also depicted for comparison.
}
\end{figure}

\begin{figure}
  \begin{center}
    \FigureFile(150mm,200mm){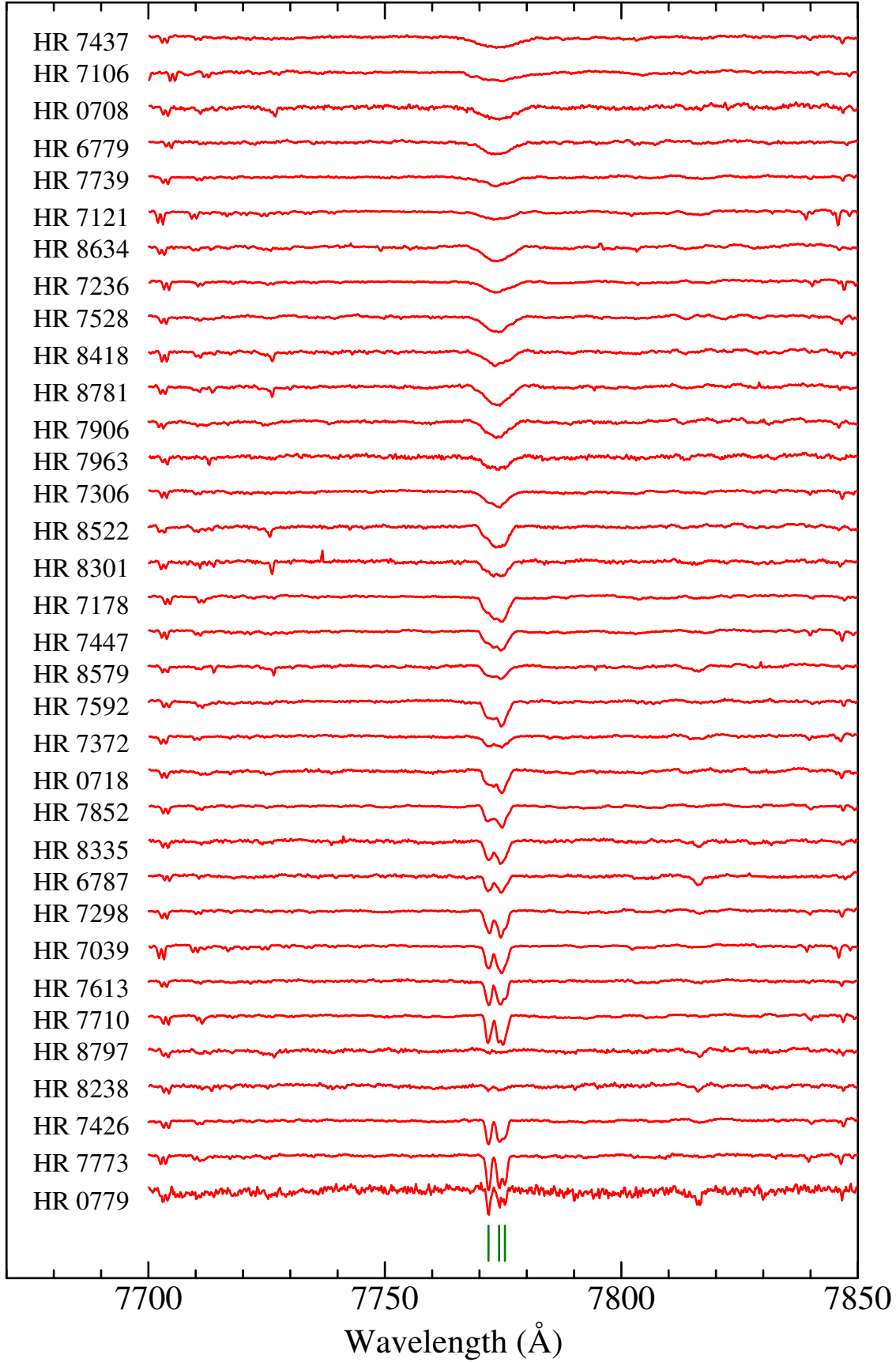}
  \end{center}
\caption{
Observed spectra of 34 B-type stars in the 7700--7850~$\rm\AA$ region
(normalized with respect to the local continuum), which are arranged 
in the descending order of $v_{\rm e}\sin i$ (from top to bottom).  
The wavelength scale is adjusted to the laboratory frame and each 
spectrum is vertically shifted by 0.2 relative to the adjacent one. 
The positions of the O~{\sc i} 7771--5 triplet lines are indicated
by vertical ticks. 
}
\end{figure}

\begin{figure}
  \begin{center}
    \FigureFile(120mm,180mm){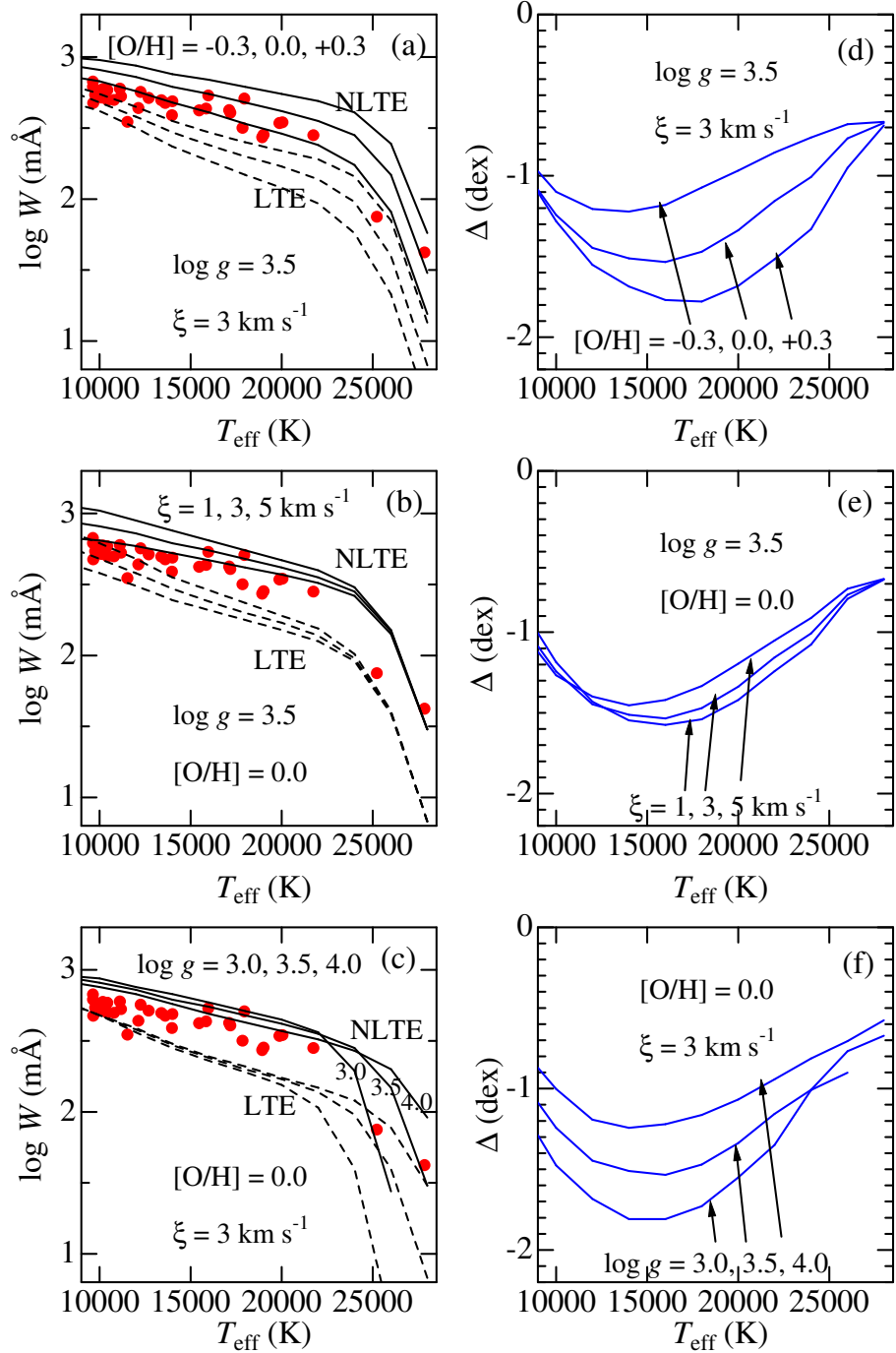}
  \end{center}
\caption{
Non-LTE (solid line) as well as LTE (dashed line) equivalent widths
($\log W$; left panels a--c) and the non-LTE abundance corrections
($\Delta$; right panels d--f) for the total O~{\sc i} 7771--5 triplet 
lines, which were theoretically computed 
for representative models and plotted against $T_{\rm eff}$.  
The upper (a,d), middle (b,e), and bottom (c,f) panels show the 
effects of changing [O/H], $\xi$, and $\log g$, respectively.
(Note that [O/H] = 0.0 corresponds to $A$ = 8.93, as remarked
in footnote~5 of section~4.) 
In the three left panels are also plotted (in filled circles) 
the $W$ values derived for our 34 targets.
}
\end{figure}

\begin{figure}
  \begin{center}
    \FigureFile(150mm,200mm){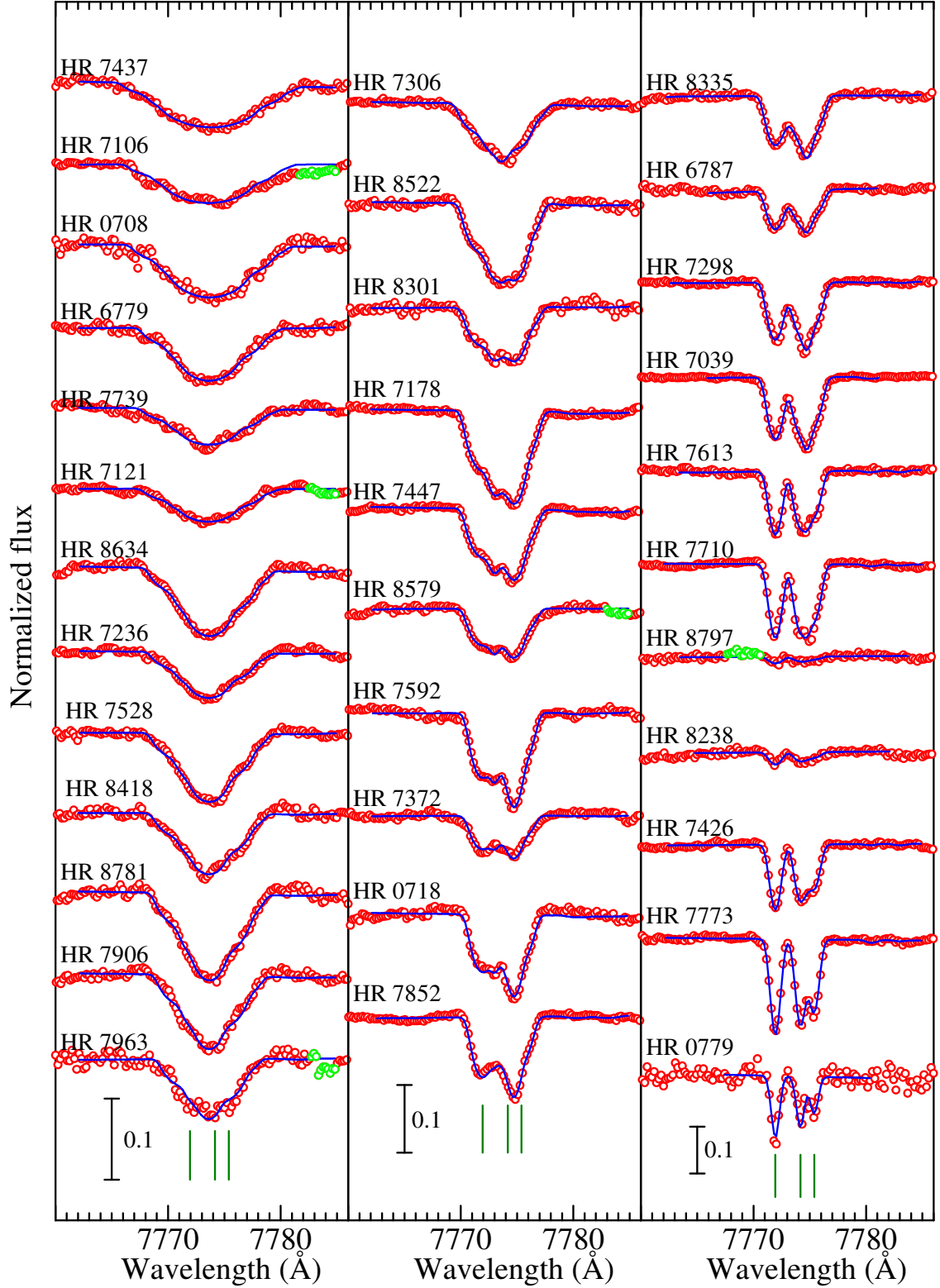}
  \end{center}
\caption{
Synthetic spectrum fitting in the region of O~{\sc i} 7771--5 
triplet lines accomplished by varying $v_{\rm e}\sin i$ and $A$(O).
The best-fit theoretical spectra are shown by blue solid lines. 
The observed data are plotted by symbols, where those actually used 
in the fitting are depicted in red, while those rejected are highlighted in green.
The spectra are arranged (from top to bottom) 
in the descending order of $v_{\rm e}\sin i$ as in table 1, 
and offset values of 0.1 (left panel), 0.15 (middle panel), 
and 0.2 (right panel) are applied to each spectrum relative to 
the adjacent one. 
}
\end{figure}

\begin{figure}
  \begin{center}
    \FigureFile(100mm,100mm){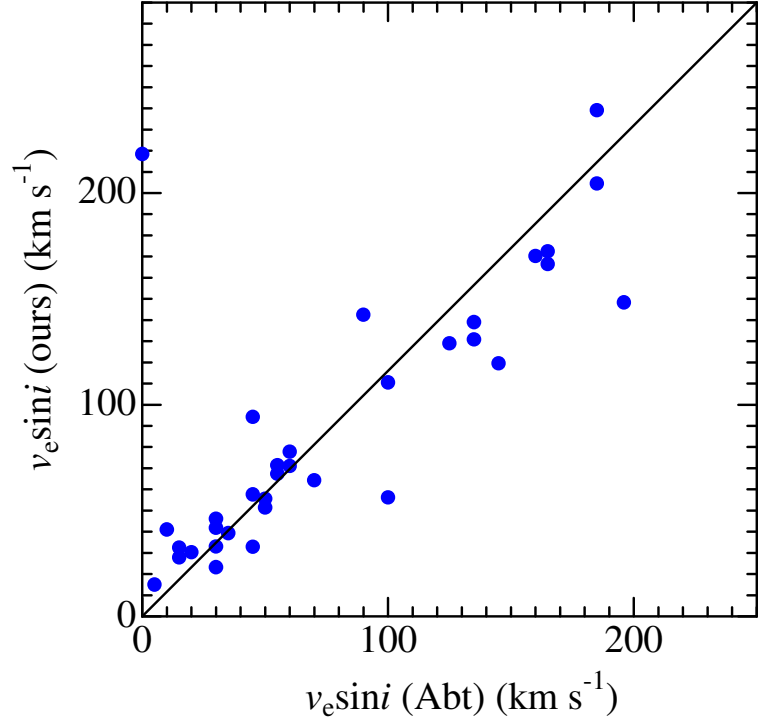}
  \end{center}
\caption{
Comparison of projected rotational velocities ($v_{\rm e}\sin i$) derived 
from our analysis with those determined by Abt et al. (2002).
}
\end{figure}

\begin{figure}
  \begin{center}
    \FigureFile(120mm,160mm){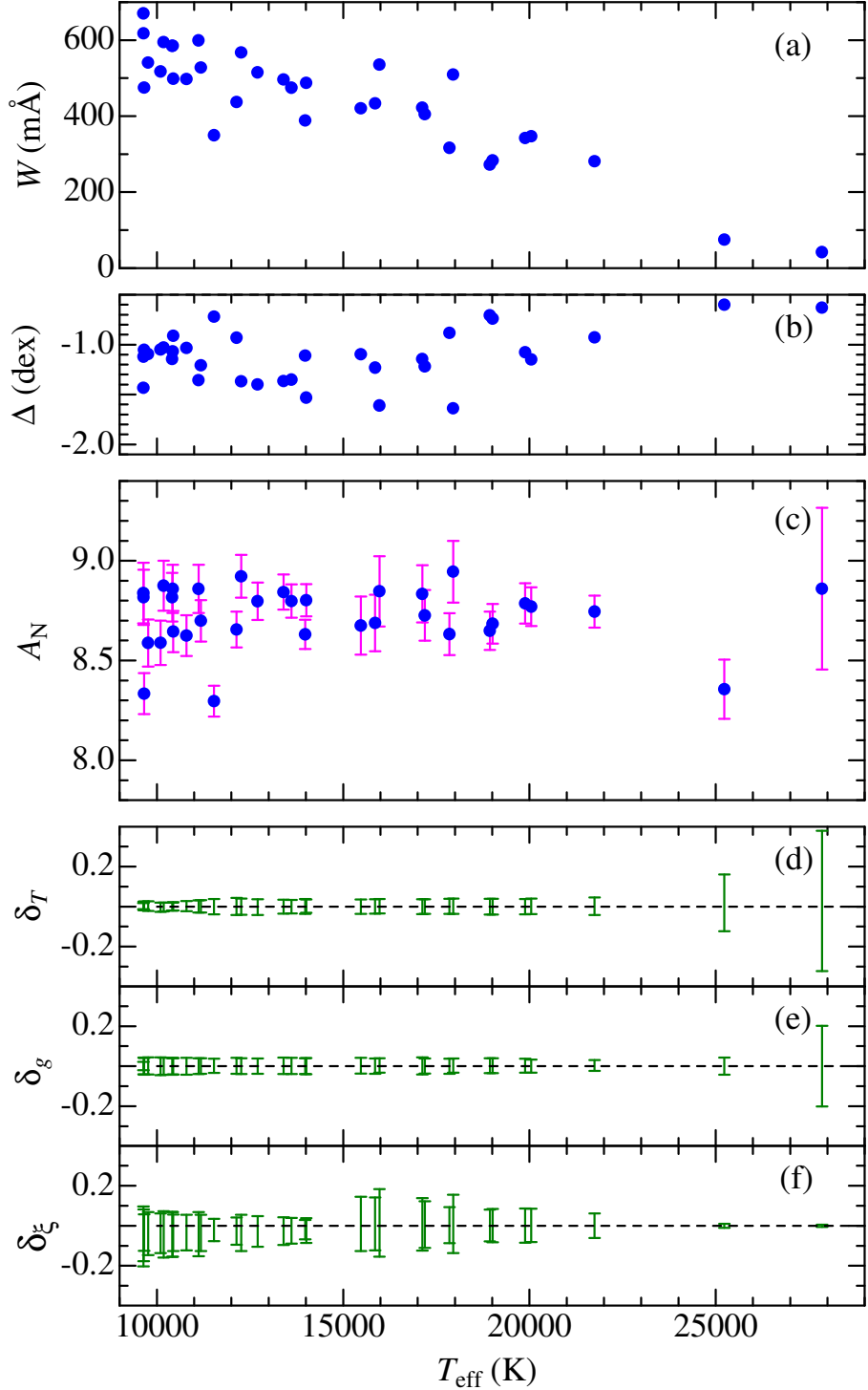}
  \end{center}
\caption{
Oxygen abundances along with the abundance-related quantities 
specific to the O~{\sc i} 7771--5 feature, plotted against $T_{\rm eff}$. 
(a) $W$ (total equivalent width of the triplet), 
(b) $\Delta$ (non-LTE correction),
(c) $A_{\rm N}$ (non-LTE oxygen abundance, where the attached 
error bar is $\delta_{Tg\xi}$), 
(d) $\delta_{T+}$ and $\delta_{T-}$ (abundance variations 
in response to $T_{\rm eff}$ changes), 
(e) $\delta_{g+}$ and $\delta_{g-}$ (abundance variations 
in response to $\log g$ changes), 
and (f) $\delta_{\xi +}$ and $\delta_{\xi -}$ (abundance 
variations in response to $\xi$ changes). See subsection~5.3
for more detailed explanations regarding the $\delta$ values.
The signs of $\delta$'s concerning the variations of $T_{\rm eff}$ and 
$\xi$ are are $\delta_{T+}>0$, $\delta_{T-}<0$, $\delta_{\xi +}<0$, and 
$\delta_{\xi -}>0$. Regarding the sensitivity to $\log g$, 
$\delta_{g+}>0$ and $\delta_{g-}<0$ for most stars, but the sign is 
reversed for the two stars at $T_{\rm eff} > 25000$~K.
}
\end{figure}

\begin{figure}
  \begin{center}
    \FigureFile(100mm,160mm){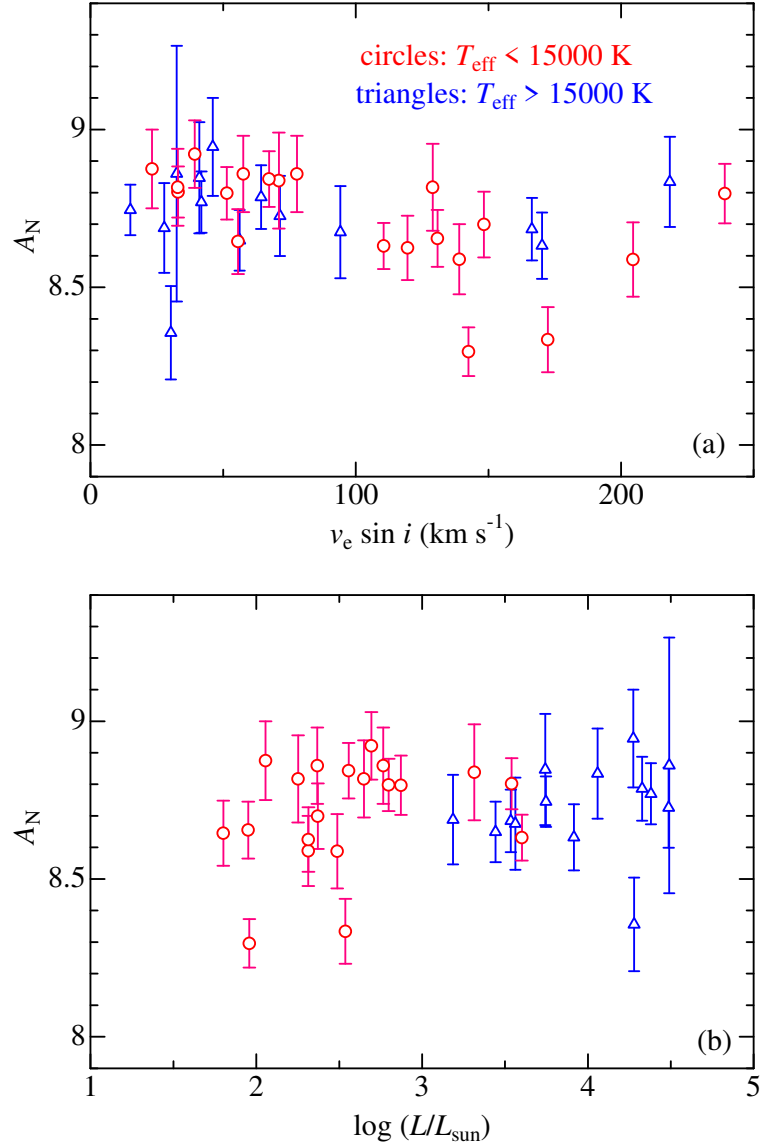}
  \end{center}
\caption{
Oxygen abundances of 34 B-type stars plotted against (a) $v_{\rm e}\sin i$
and (b) $\log L$ (cf. figure 1).
where stars of $T_{\rm eff} < 15000$~K and $T_{\rm eff} > 15000$~K,
are expressed in open circles and open triangles, respectively. 
The data and the attached error bars are the same as in figure~6c. 
}
\end{figure}

\setcounter{table}{0}
\begin{table}[h]
\scriptsize
\caption{Basic data of the program stars and the results of the analysis.}
\begin{center}
\begin{tabular}
{ccccc rcrc rccc}\hline \hline
HR\# & HD\# & Desig. & Sp.type & $V$ & $T_{\rm eff}$ & $\log g$ & $v_{\rm e} \sin i$ &
$\xi$ & $W$ & $A_{\rm N}$ & $\Delta$ & Obs. date \\
\hline
HR~7437 &HD~184606 &  9~Vul          &B8~IIIn    & 5.00 & 12703 & 3.52 & 239.1 &  1 & 515 & 8.80 &$-$1.40 & 150731\\
HR~7106 &HD~174638 &  $\beta$~Lyr    &B8.5~Ib-II & 3.52 & 17124 & 3.92 & 218.4 &  3 & 422 & 8.83 &$-$1.14 & 150612\\
HR~0708 &HD~015130 &  $\rho$~Cet     &A0~V       & 4.88 &  9764 & 3.48 & 204.5 &  1 & 541 & 8.59 &$-$1.10 & 150926\\
HR~6779 &HD~166014 &  $o$~Her &B9.5~III   & 3.84 &  9657 & 3.24 & 172.4 &  1 & 475 & 8.33 &$-$1.05 & 150501\\
HR~7739 &HD~192685 &  $\cdots$       &B3~V       & 4.79 & 17855 & 3.94 & 170.2 &  3 & 317 & 8.63 &$-$0.88 & 150815\\
HR~7121 &HD~175191 &  $\sigma$~Sgr   &B2~V       & 2.05 & 19012 & 4.27 & 166.4 &  3 & 283 & 8.68 &$-$0.74 & 150731\\
HR~8634 &HD~214923 &  $\zeta$~Peg    &B8~V       & 3.41 & 11182 & 3.65 & 148.3 &  1 & 528 & 8.70 &$-$1.21 & 150927\\
HR~7236 &HD~177756 &  $\lambda$~Aql  &B9~Vn      & 3.43 & 11534 & 4.14 & 142.5 &  1 & 350 & 8.30 &$-$0.72 & 150731\\
HR~7528 &HD~186882 &  $\delta$~Cyg   &B9.5~IV+.  & 2.86 & 10098 & 3.67 & 139.0 &  1 & 518 & 8.59 &$-$1.05 & 150926\\
HR~8418 &HD~209819 &  $\iota$~Aqr    &B8~V       & 4.29 & 12138 & 4.26 & 130.8 &  1 & 437 & 8.65 &$-$0.93 & 150927\\
HR~8781 &HD~218045 &  $\alpha$~Peg   &B9~III     & 2.49 &  9643 & 3.52 & 129.0 &  1 & 618 & 8.82 &$-$1.12 & 150927\\
HR~7906 &HD~196867 &  $\alpha$~Del   &B9~IV      & 3.77 & 10794 & 3.89 & 119.5 &  1 & 497 & 8.62 &$-$1.03 & 150926\\
HR~7963 &HD~198183 &  $\lambda$~Cyg  &B5~V       & 4.53 & 13982 & 3.89 & 110.5 &  1 & 389 & 8.63 &$-$1.11 & 150927\\
HR~7306 &HD~180554 &  1~Vul          &B4~IV      & 4.76 & 15474 & 3.78 &  94.2 &  3 & 421 & 8.68 &$-$1.10 & 150731\\
HR~8522 &HD~212097 &  32~Peg         &B9~III     & 4.78 & 11119 & 3.44 &  77.8 &  1 & 599 & 8.86 &$-$1.36 & 150927\\
HR~8301 &HD~206672 &  $\pi^{1}$~Cyg  &B3~III     & 4.69 & 17191 & 3.55 &  71.4 &  3 & 405 & 8.73 &$-$1.22 & 150927\\
HR~7178 &HD~176437 &  $\gamma$~Lyr   &B9~III     & 3.25 &  9638 & 2.73 &  71.0 &  1 & 670 & 8.84 &$-$1.43 & 150612\\
HR~7447 &HD~184930 &  $\iota$~Aql    &B5~III     & 4.36 & 13400 & 3.69 &  67.3 &  1 & 497 & 8.84 &$-$1.37 & 150731\\
HR~8579 &HD~213420 &  6~Lac          &B2~IV      & 4.52 & 19885 & 3.67 &  64.3 &  3 & 343 & 8.79 &$-$1.08 & 150927\\
HR~7592 &HD~188260 &  13~Vul         &B9.5~III   & 4.57 & 10425 & 3.89 &  57.6 &  1 & 585 & 8.86 &$-$1.07 & 150815\\
HR~7372 &HD~182568 &  2~Cyg          &B3~IV      & 4.99 & 18937 & 4.28 &  56.2 &  3 & 272 & 8.65 &$-$0.71 & 150801\\
HR~0718 &HD~015318 &  $\xi^{2}$~Cet  &B9~III     & 4.30 & 10438 & 4.09 &  55.6 &  1 & 498 & 8.65 &$-$0.91 & 150926\\
HR~7852 &HD~195810 &  $\epsilon$~Del &B6~III     & 4.03 & 13611 & 3.68 &  51.4 &  1 & 475 & 8.80 &$-$1.35 & 150815\\
HR~8335 &HD~207330 &  $\pi^{2}$~Cyg  &B3~III     & 4.23 & 17953 & 3.23 &  46.1 &  3 & 509 & 8.94 &$-$1.64 & 150927\\
HR~6787 &HD~166182 &  102~Her        &B2~IV      & 4.37 & 20048 & 3.45 &  41.8 &  3 & 347 & 8.77 &$-$1.15 & 150501\\
HR~7298 &HD~180163 &  $\eta$~Lyr     &B2.5~IV    & 4.43 & 15974 & 3.19 &  41.0 &  3 & 535 & 8.85 &$-$1.61 & 150801\\
HR~7039 &HD~173300 &  $\phi$~Sgr     &B8~III     & 3.17 & 12263 & 3.66 &  39.3 &  1 & 567 & 8.92 &$-$1.37 & 150815\\
HR~7613 &HD~188892 &  22~Cyg         &B5~IV      & 4.95 & 14008 & 3.38 &  33.0 &  1 & 487 & 8.80 &$-$1.53 & 150815\\
HR~7710 &HD~191692 &  $\theta$~Aql   &B9.5~III   & 3.24 & 10410 & 3.71 &  32.9 &  1 & 585 & 8.82 &$-$1.14 & 150815\\
HR~8797 &HD~218376 &  1~Cas          &B0.5~III   & 4.84 & 27853 & 3.64 &  32.5 &  3 &  42 & 8.86 &$-$0.63 & 150927\\
HR~8238 &HD~205021 &  $\beta$~Cep    &B0.5~IIIs  & 3.23 & 25231 & 3.63 &  30.3 &  3 &  75 & 8.36 &$-$0.60 & 150927\\
HR~7426 &HD~184171 &  8~Cyg          &B3~IV      & 4.74 & 15858 & 3.54 &  27.8 &  3 & 434 & 8.69 &$-$1.23 & 150801\\
HR~7773 &HD~193432 &  $\nu$~Cap      &B9~IV      & 4.77 & 10180 & 3.91 &  23.2 &  1 & 595 & 8.88 &$-$1.03 & 150801\\
HR~0779 &HD~016582 &  $\delta$~Cet   &B2~IV      & 4.08 & 21747 & 3.63 &  15.0 &  3 & 281 & 8.74 &$-$0.93 & 150926\\
\hline
\end{tabular}
\end{center}
Note. \\
In columns 1 through 8 are given the HR number, HD number, star designation, 
spectral type, apparent visual magnitude (in mag), 
effective temperature (in K), logarithmic surface gravity (in cm~s$^{-2}$),
and  projected rotational velocity (in km~s$^{-1}$; derived from our 
fitting analysis). Columns 9 through 12 show the results of abundance analysis:
assigned microturbulence (cf. section~3), total equivalent width of the 
O~{\sc i}~7771--5 lines (in m$\rm\AA$), non-LTE O abundance (in dex; expressed 
in the usual normalization of H=12.00), and non-LTE correction (in dex). 
Column 13 gives the observation date (for example, 150907 means 2015 September 7). 
The objects are arranged in the descending order of $v_{\rm e} \sin i$. 
\end{table}

\setcounter{table}{1}
\footnotesize
\begin{table}[h]
\caption{Adopted atomic data of O~{\sc i} 7771--5 lines.}
\begin{center}
\begin{tabular}{ccccccc}\hline\hline
Species & $\lambda_{\rm air}$ & $\chi_{\rm low}$ & $\log gf$ & Gammar & Gammas &
Gammaw \\
        & ($\rm\AA$) & (eV) & (dex) & (dex) & (dex) & (dex) \\
\hline
O~{\sc i} & 7771.944 & 9.146&+0.324 & 7.52 & $-$5.55 & ($-$7.65) \\
O~{\sc i} & 7774.166 & 9.146&+0.174 & 7.52 & $-$5.55 & ($-$7.65) \\
O~{\sc i} & 7775.388 & 9.146&$-$0.046 & 7.52 & $-$5.55 & ($-$7.65) \\
\hline
\end{tabular}
\end{center}
\footnotesize
Note.\\
The first four columns are self-explanatory.
Presented in columns 5--7 are the damping parameters: 
Gammar is the radiation damping width (s$^{-1}$) [$\log\gamma_{\rm rad}$], 
Gammas is the Stark damping width (s$^{-1}$) per electron density (cm$^{-3}$) 
at $10^{4}$ K [$\log(\gamma_{\rm e}/N_{\rm e})$], and
Gammaw is the van der Waals damping width (s$^{-1}$) per hydrogen density 
(cm$^{-3}$) at $10^{4}$ K [$\log(\gamma_{\rm w}/N_{\rm H})$]. \\
These data were taken from the compilation of Kurucz and Bell (1995)
as long as available, while the parenthesized damping parameters are 
the default values computed by the WIDTH9 program.
\end{table}


\begin{thebibliography}{}
\bibitem[]{}
  Abt, A. A., Levato, H., \& Grosso, M. 2002, ApJ, 573, 359
\bibitem[]{} 
  Anders, E., \& Grevesse, N. 1989, Geochim. Cosmochim. Acta, 53, 197
\bibitem[]{}
  Arenou, F., Grenon, M., \& G\'{o}mez, A. 1992, A\&A, 258, 104
\bibitem[]{}
  Asplund, M., Grevesse, N., Sauval, A. J., \& Scott, P. 2009,
  ARA\&A, 47, 481
\bibitem[]{}
  Cunha, K., \& Lambert, D. L. 1994, ApJ, 426, 170
\bibitem[]{}
  ESA 1997, The Hipparcos and Tycho Catalogues, ESA SP-1200, 
  available from NASA-ADC or CDS in a machine-readable form 
  (file name: hip\_main.dat)
\bibitem[]{} 
  Flower, P. J. 1996, ApJ, 469, 355
\bibitem[]{}
  Girardi, L., Bressan, A., Bertelli, G., \& Chiosi, C. 2000,
  A\&AS, 141, 371
\bibitem[]{}
  Georgy, C., Ekstr\"{o}m, S., Granada, A., Meynet, G., Mowlavi, N.,
  Eggenberger, P., \& Maeder, A. 2013, A\&A, 553, A24
\bibitem[]{}
  Gies, D. R., \& Lambert, D. L. 1992, ApJ, 387, 673
\bibitem[]{}
  Gray, D. F. 2005, The Observation and Analysis of Stellar Photospheres, 3rd ed.
  (Cambridge: Cambridge University Press)
\bibitem[]{}
  Hauck, B., \& Mermilliod, M. 1998, A\&AS, 129, 431
\bibitem[]{}
  Hempel, M., \& Holweger, H. 2003, A\&A, 408, 1065
\bibitem[]{}
  Kilian, J. 1992, A\&A, 262, 171
\bibitem[]{}
  Korotin, S. A., Andrievsky, S. M., \& Luck, R. E. 1999, A\&A, 351, 168
\bibitem[]{}
  Kurucz, R. L. 1993, Kurucz CD-ROM, No. 13 (Harvard-Smithsonian Center
  for Astrophysics)
\bibitem[]{}
  Kurucz, R. L., \& Bell, B. 1995, Kurucz CD-ROM, No. 23 
  (Harvard-Smithsonian Center for Astrophysics)
\bibitem[]{}
  Lyubimkov, L. S.,  Lambert, D. L. , Poklad, D. B., Rachkovskaya, T. M., 
  \& Rostopchin, S. I. 2013, MNRAS, 428, 3497
\bibitem[]{}
  Lyubimkov, L. S., Rostopchin, S. I., \& Lambert, D. L. 2004, 
  MNRAS, 351, 745
\bibitem[]{}
  Napiwotzki, R., Sch\"{o}nberner, D., \& Wenske, V. 1993,
  A\&A, 268, 653
\bibitem[]{}
  Niemczura, E., Morel, T., Aerts, C. 2009, A\&A, 506, 213
\bibitem[]{}
  Nieva, M.-F. 2013, A\&A, 550, A26
\bibitem[]{}
  Nieva, M.-F., \& Przybilla, N. 2012, A\&A, 539, A143
\bibitem[]{}
  Ozaki, S., \& Tokimasa, N. 2005, Annu. Rep. Nishi-Harima Astron. Obs., 
  15, 15 (in Japanese)
\bibitem[]{}
  Sim\'{o}n-D\'{\i}az, S. 2010, A\&A, 510, A22
\bibitem[]{}
Takeda, Y. 1991, A\&A, 242, 455
\bibitem[]{}
  Takeda, Y. 1992, PASJ, 44, 309
\bibitem[]{}
  Takeda, Y. 1995, PASJ, 47, 287
\bibitem[]{}
  Takeda, Y. 2003, A\&A, 402, 343
\bibitem[]{}
  Takeda, Y., Han, I., Kang, D.-I., Lee, B.-C., \& Kim, K.-M. 2008,
  JKAS, 41, 83
\bibitem[]{}
  Takeda, Y., \& Honda, S. 2005, PASJ, 57, 65
\bibitem[]{}
  Takeda, Y., \& Honda, S. 2015, PASJ, 67, 25
\bibitem[]{}
  Takeda, Y., Kambe, E., Sadakane, K., \& Masuda, S. 2010. PASJ, 62, 1239 (Paper~I)
\bibitem[]{}
  Takeda, Y., Kawanomoto, S., \& Ohishi, N. 2014, PASJ, 66, 23
\bibitem[]{}
  Takeda, Y., Kawanomoto, S., \& Sadakane, K. 1998, PASJ, 50, 97
\bibitem[]{}
  Takeda, Y., \& Sadakane, K. 1997, PASJ, 49, 367
\bibitem[]{}
  Takeda, Y., Sato, B., Omiya, M., \& Harakawa, H. 2015, PASJ, 67, 24
\end{thebibliography}
\end{document}